\documentclass[A4]{aa}
\usepackage{graphicx}
\def\erg{erg cm$^{-2}$ s$^{-1}$}
\def\xmm{{\it XMM-Newton~}}
\def \nh {N${\rm _H}$~}

\newcommand {\wga} {{1WGA~J2223.7-0206~}}
\begin{document}
\title{1WGA~J2223.7-0206: a Narrow--Line Quasi--Stellar Object in the
XMM-Newton field of view of 3C445\thanks{Based on observations obtained 
with \emph{XMM-Newton}, an ESA science mission with instruments and
contributions directly funded by ESA Member States and the USA (NASA),
with the Bologna Astronomical Observatory in Loiano, Italy, and with
the WHT telescope of the ING at La Palma, Spain.}
}

\author{Paola Grandi, Luigi Foschini, Nicola Masetti, Eliana Palazzi}

\offprints{P. Grandi \email{grandi@bo.iasf.cnr.it}}

\institute{Istituto di Astrofisica Spaziale e Fisica Cosmica (IASF) del
CNR, Sezione di Bologna, Via Gobetti 101, 40129 Bologna (Italy)}

\date{Received 2003 November 21; accepted 2004 January 30}

\abstract{We report the discovery of a Narrow Line QSO located at
about $1.3'$ from the Broad Line Radio Galaxy 3C445. The source, \wga, although already revealed by ROSAT, 
has never been optically identified previously. An \xmm observation of 3C445 has allowed, for the first time, an accurate  X-ray spectral study of \wga, revealing 
an ultra-soft spectrum and fast flux variations typical of Narrow Line AGN.
The 0.2-10 keV spectrum is well represented by a power law ($\Gamma=2.5$)
plus a black body component ($kT = 117$~eV) absorbed by Galactic \nh.
About $80\%$ of the X-ray flux ($F_{0.2-10~\rm keV} \sim
3\times10^{-13}$~\erg) is emitted below $2$~keV.
The $0.2-2$~keV flux is observed to decrease by about a factor $1.6$ in about $5000$~s. 

The optical observations, triggered by the X-ray study,  confirm the Narrow Line AGN nature of this source. 
The continuum is blue with typical AGN emission lines, pointing to a redshift $z=0.46$. 
The full width half maximum of H$_\beta$  is $2000$~km s$^{-1}$ and the flux ratio [O {\sc iii}]/H$\beta=0.21$.
The optical luminosity  ($M_R=-23.2$) and the point-like appearance in the optical images identify \wga 
as a Narrow Line QSO. 

From the optical--UV--X--ray Spectral Energy Distribution we obtain 
a lower limit of the bolometric luminosity of \wga ($L_{\rm bol} \ge 3\times
10^{45}$ erg s$^{-1}$) implying, for accretion rates close to the Eddington limit, a
black hole mass $M_{\rm BH} \ge 2.4 \times10^{7}$~$M_{\odot}$.
}

\authorrunning{P. Grandi et al.}
\titlerunning{Narrow--line QSO in the FOV of 3C445}

\maketitle

\section{Introduction}

The Narrow Line (NL) Type I AGN (Seyfert and QSO) are a class of objects
which has focused the attention of the scientific community in the last 
years because of its unusual optical and X-ray properties. Their optical
spectra have permitted lines which are slightly broader than the forbidden
lines (i.e.full width at half maximum (FWHMH) H$_\beta\le 2000$ km $s^{-1}$), 
a flux ratio O[{\sc III}]$\lambda5007/H_{\beta}<3$ 
and a prominent Fe{\sc II} bump (Osterbrock $\&$ Pogge 1985). In the X-ray range, they exhibit strong flux
variability and very steep spectra (Boller et al. 1996, Forster \& Halpern
1996, Molthagen et al. 1998, Dewangan et al. 2001). The most favored
interpretation is that they represent an AGN class with very small black
holes and very high accretion rates.  
Actually, emission lines of highly ionized iron,
as expected to be produced by high accreting disks (Nayakshin \& Kazanas
2001), have been discovered in several NL Seyfert 1 galaxies with ASCA and
BeppoSAX (Pounds et al. 1995; Comastri et al. 1998; Turner et al. 1998;
Leighly 1999, Gliozzi et al. 2001).
Recently, Pounds et al. (2003a,b) have revealed several absorption
lines in both \xmm high and low resolution spectra, interpreted as signatures of 
a highly ionized wind. As suggested by King $\&$ Pounds (2003) 
AGN with very high accretion rates should be able to produce an outflowing photosphere.

In this paper, we present  \xmm and optical observations of the newly 
discovered NLQSO located only $1.3'$ far away from the Broad Line Radio Galaxy 3C445. 
This serendipitous source is included in the the WGA catalog (White, Giommi, Angelini 1994) with the name \wga and
discussed in a ROSAT 3C445 study of Sambruna et al. (1998).
However no particular attention has been  paid to it. Its
faintness in the ROSAT band (about a factor 3 below 3C445) assured a negligible contamination of the AGN spectrum.

Surprisingly enough, when \xmm pointed the same field, the serendipitous
source, this time clearly spatially resolved, appeared as bright as 3C445 in the
soft energy band ($0.2-1$~keV). As we will show in this paper, optical 
observations of \wga indicate that the source is actually a quasar at 
$z=0.46$ with spectral features typical of NL Type I AGN.
The presence of this peculiar AGN in the vicinity of 3C445 should be taken 
into account when the X-ray spectra from previous and very poor spatial resolution 
satellites are interpreted.

\begin{figure*}[!ht]
\centering
\includegraphics[scale=0.37]{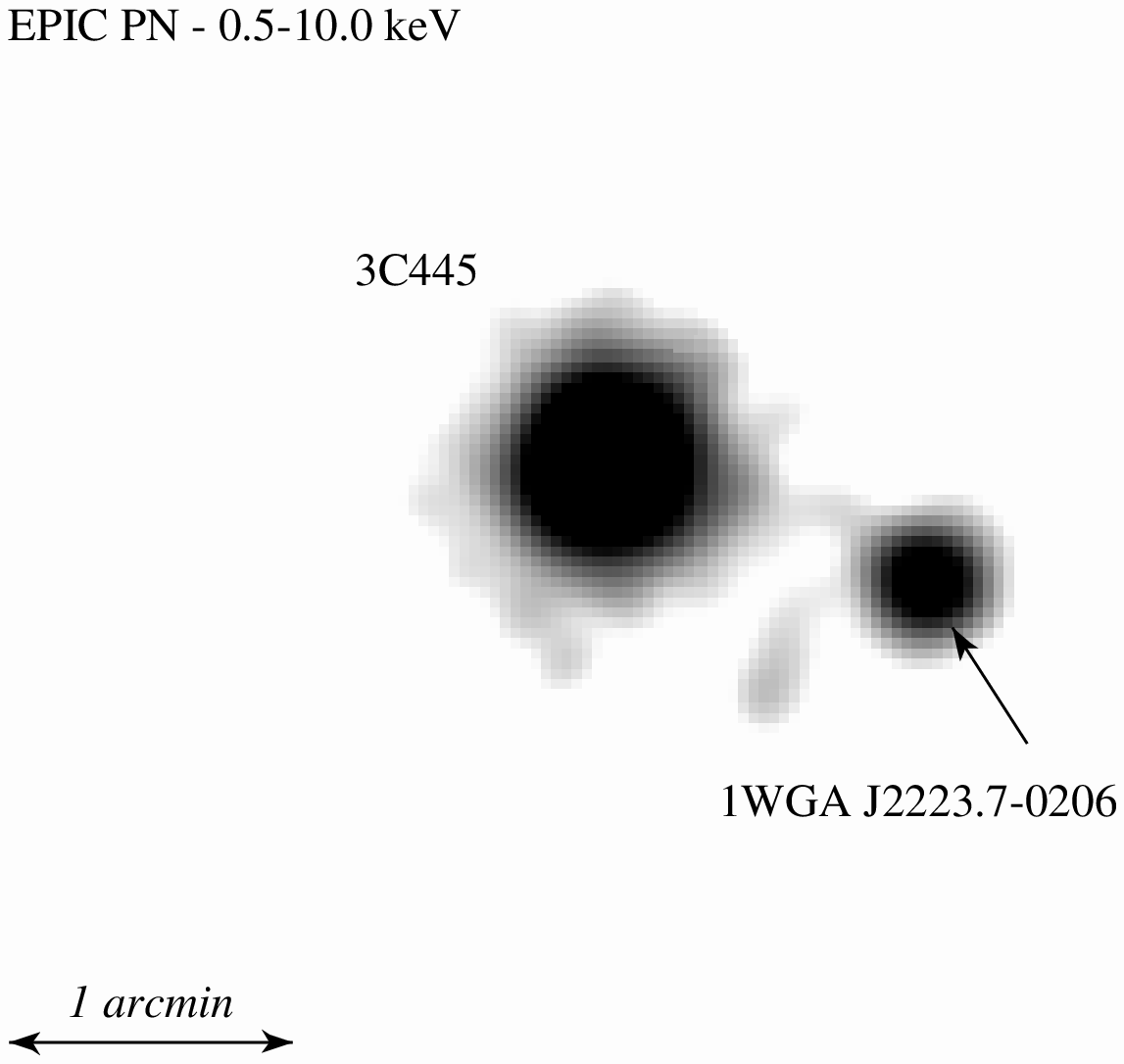}
\includegraphics[scale=0.37]{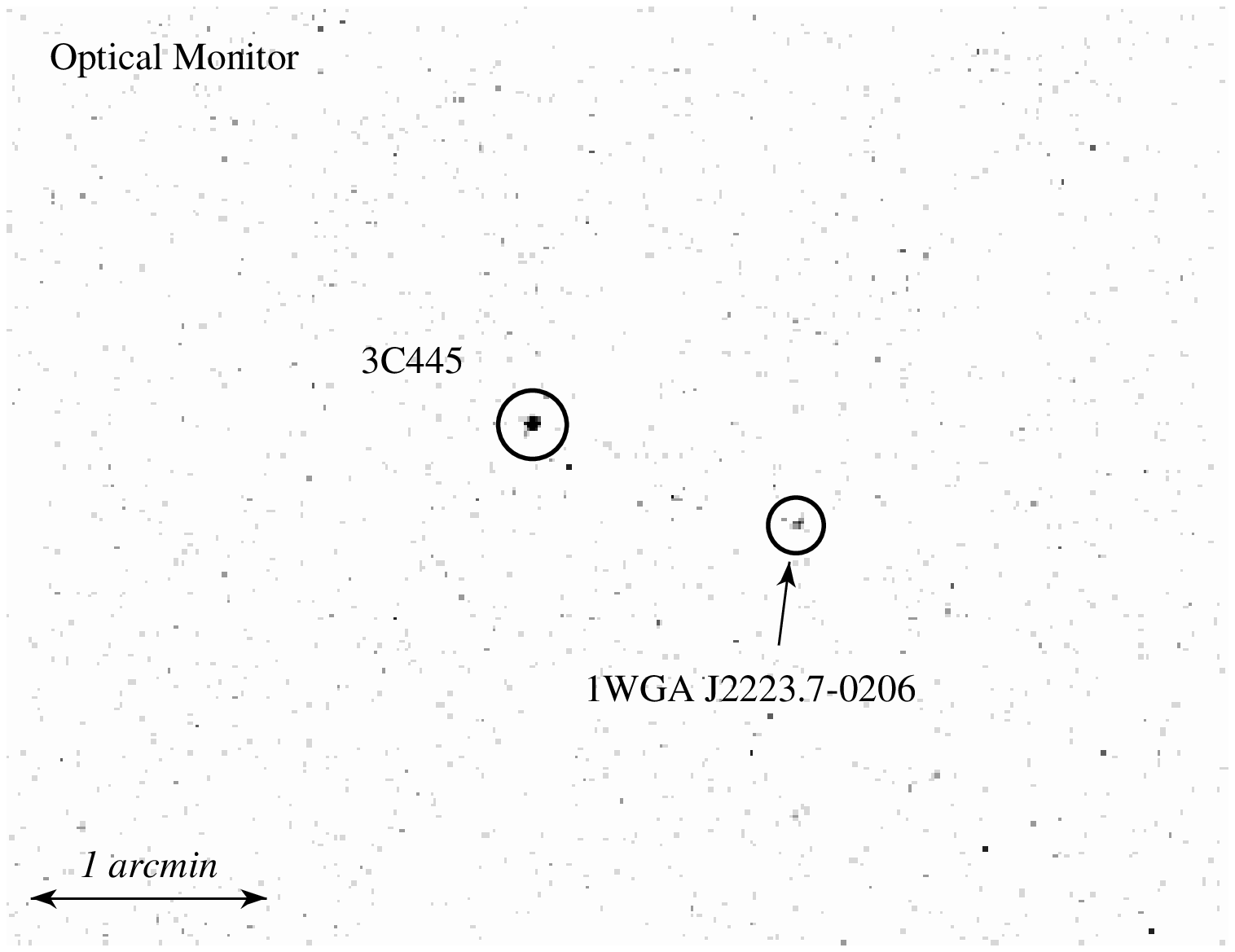}
\includegraphics[scale=0.37]{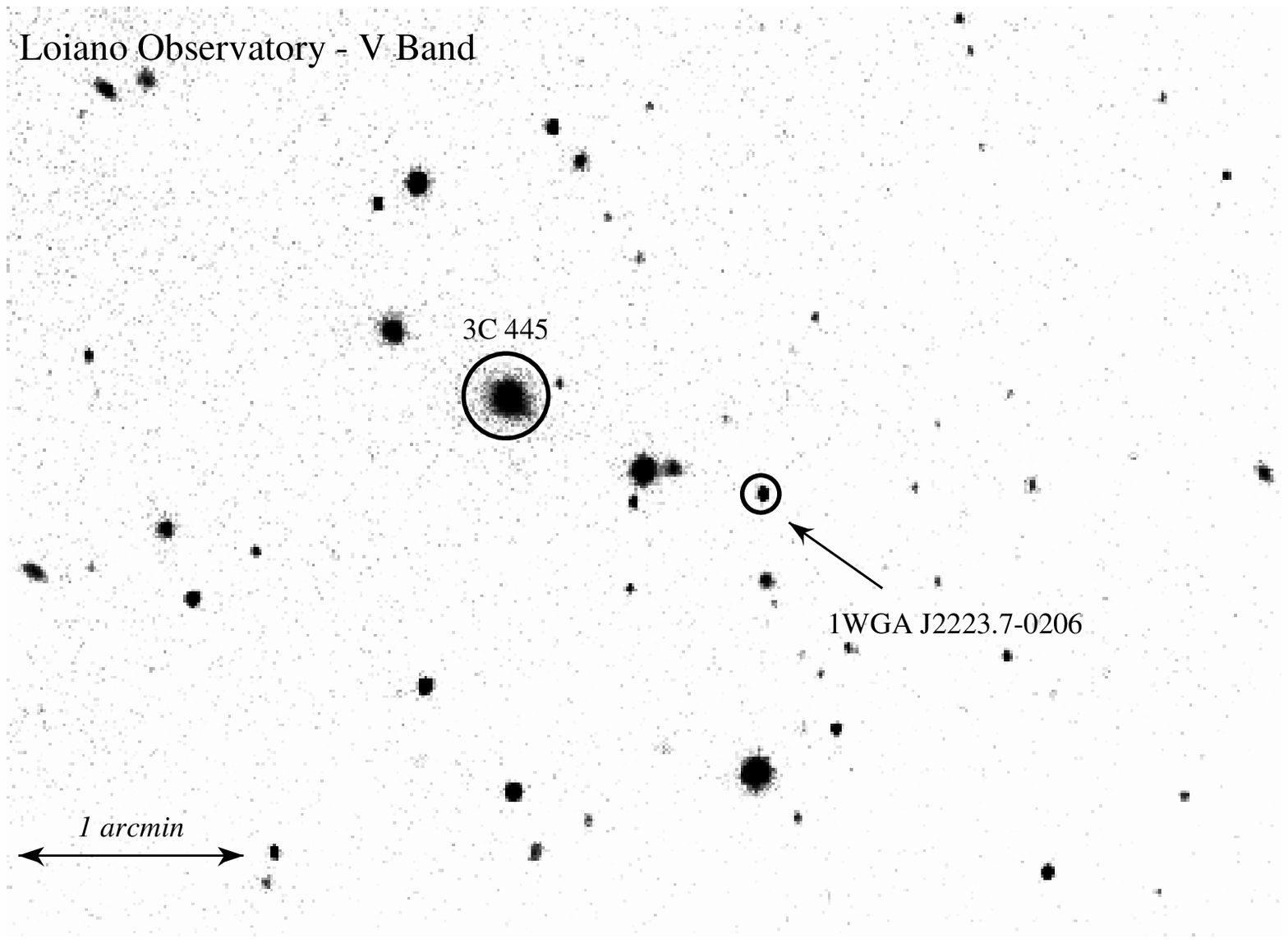}
\caption{EPIC-PN ({\it  left panel}), OM/UVW2 ({\it  middle  panel}) and 
Loiano Observatory  V band ({\it right panel}) images of \wga. 
The X-ray and Ultraviolet images are smoothed using a Gaussian filter
with $\sigma=2$~pixels (corresponding to $8''$ and $1''$ in the PN and OM, 
respectively).}
\label{om}
\end{figure*}

Throughout the paper, luminosities are calculated assuming
isotropic emission, a Hubble constant of $H_0=75$~km~s$^{-1}$~Mpc$^{-1}$
and a deceleration parameter of $q_0 = 0.5$.

\section{\xmm Observation}

\emph{XMM--Newton} observed 3C445 on 2001 December 6 with the EPIC cameras MOS (Turner et al. 2001) and
PN (Str\"uder et al. 2001) in small window mode for 20~ks.

A serendipitous source was detected in the PN image at coordinates
$\alpha=22:23:44.9$, $\delta=-02:06:40$ (J2000, position accuracy $\sim
4''$)  at $\approx1.3'$ from 3C445 (see Figure~\ref{om}, left  panel). It
was not revealed by the MOS cameras, because of the smaller field of view.
A rapid check of the WGA catalog allowed us to immediately identify it with
the source \wga, detected by ROSAT at coordinates $\alpha=22:23:44.7$,
$\delta=-02:06:33$ (J2000, WGA position accuracy $\sim 13''$).

Data from the Optical Monitor (OM, Mason et al. 2001) were also available
(Figure~\ref{om}, middle panel). 
We got the mosaic image in the bands UVW2
($180-225$~nm) and UVM2 ($205-245$~nm) from the standard pipeline.
WGA~J$2223.7-0206$ was detected in both bands at the
coordinates $\alpha=22:23:45.10$ and $\delta=-02:06:41$ (J2000, $0.9''$ of
error radius) with significance $14\sigma$ in UVW2 and $19\sigma$ in UVM2.  
UV magnitudes of $m_{\mathrm{UVW2}}=17.4\pm 0.1$ and
$m_{\mathrm{UVM2}}=17.8\pm 0.1$ were obtained following the recent
prescriptions of Chen (2003).

\subsection{X-ray timing and spectral analysis}

For the processing and screening of the data we followed the standard procedures described in Snowden et al. (2002). 
The XMM--SAS (5.4.1) software was utilized. Spectral and time analysis were performed with Xspec (11.2.0) and Xronos (5.19).
 
The spectrum and the light curve of the source were extracted
using a circle of $15''$ radius to take into account any  possible 
3C445 contamination. The background was estimated in an annular
region of inner circle radius $15''$ and width $25''$\footnote{We also extracted a background spectrum from a region free 
of sources using a circle of $40''$ radius. The spectral fit results were completely consistent with those obtained using a background spectrum 
from an annulus surrounding the source.}. 

The spectrum was rebinned so that each energy bin contained a minimum of 25
photons and it was fitted in the $0.2-10$~keV energy range. The appropriate  photon
redistribution matrix and ancillary file were created with the \texttt{rmfgen} and \texttt{arfgen} tasks of XMM--SAS.

Fast variability occurred during the observation.  The source was in a
higher activity state during the first $5000$~s of the observation and then it
decreased of about $40\%$ (Fig.~2). A standard $\chi^2$ test assures
that constant emission can be ruled out at a level of confidence $>99.99$\%.

\begin{figure}[h]
\centering
\includegraphics[angle=270,scale=0.35]{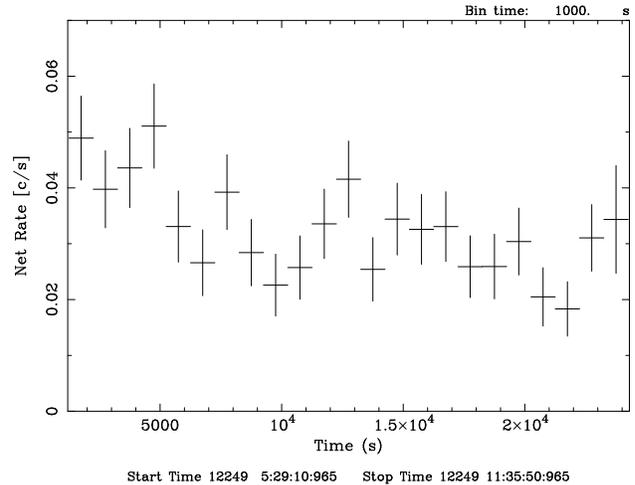}
\caption{EPIC--PN light-curve of \wga in the energy band $0.2-2$~keV.
Background subtracted source count in $1000$~s bin and associated 
1$\sigma$ errors are shown}
\label{lcr}
\end{figure}

As the two spectra, extracted before and after the flux drop, did not
display any significant change of shape, we performed the spectral analysis
on the entire observation.

\begin{figure}[h]
\centering
\includegraphics[angle=270,scale=0.35]{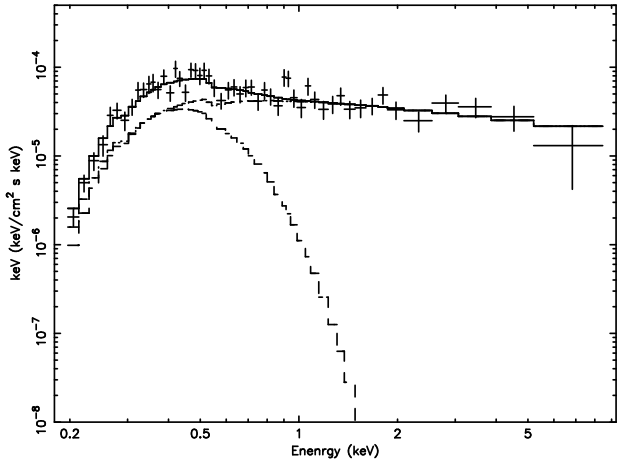}
\caption{The EPIC--PN observed spectrum of \wga.
The best fit is a power law plus a blackbody model 
absorbed by Galactic column density.}
\label{spec1}
\end{figure}

A simple absorbed power law did not give a satisfactory fit to the data. The
fit was poor ($\chi^2=62$ for 49 degrees of freedom (d.o.f.)) and an
inspection of the residuals showed an excess of emission at low energy.
When a black body model was added, the $\chi^2$ value became better
($\chi^2=47$ for  47 d.o.f.) implying a significant improvement (P$_{\rm
F_{test}}=99.2\%$).  As the acceptable range of \nh was consistent with the
Galactic line of sight value ($N_{H}=5.0\times 10^{20}$~cm$^{-2}$; Dickey
\& Lockman 1990), we fixed the column density to the Galactic value in
order to reduce the fit uncertainties. The best fit parameters are listed
in Table 1. The source is extremely soft with $\sim 80\%$ of the
radiation emitted below $2$~keV (see also Fig.~\ref{spec1}). The
intrinsic luminosity (i.e. corrected for the absorption column along the
line of sight)  in the rest frame is $2.3\times 10^{44}$~erg s$^{-1}$  in the $0.2-10$~keV band
(see below for the distance determination).

\begin{table*}[t!]
\label{xmm}
\caption{EPIC--PN best fit spectral (rest frame) parameters. Column
density was fixed to the Galactic value $N_H^{\rm Gal} = 5.0\times10^{20}$ 
cm$^{-2}$.}
\begin{center}
\vspace{-.3cm}
\begin{tabular}{ccccc}
\noalign{\smallskip}
\hline
\noalign{\smallskip}
\multicolumn{1}{c}{$\Gamma$} 
&\multicolumn{1}{c}{kT} 
&\multicolumn{1}{c}{$f^a_{\rm PL}$} 
&\multicolumn{1}{c}{$f^a_{\rm PL}$}
&\multicolumn{1}{c}{$f^a_{\rm BB}$}\\ 
&\multicolumn{1}{c}{[eV]} 
&\multicolumn{1}{c}{[$2-10$~keV]} 
&\multicolumn{1}{c}{[$0.2-2$~keV]}
&\multicolumn{1}{c}{[$0.2-2$~keV]}\\ 
\noalign{\smallskip}
\hline
\noalign{\smallskip}

2.4$\pm0.2$ &  117$\pm21$ & $0.7\pm0.1$ & $2.2\pm0.4$ & $1.1\pm0.5$ \\

\noalign{\smallskip}
\hline
\noalign{\smallskip}
\multicolumn{5}{l}{$^a$ in units of $10^{-13}$ \erg} \\
\noalign{\smallskip}
\end{tabular}
\end{center}
\end{table*}

\section{Optical observations}

The inspection of the Red Digitized Sky Survey II\footnote{available at \texttt{http://archive.eso.org/dss/dss/}}
shows, within the \xmm error box for \wga, one single and relatively bright
object. Indeed this source is also present in the USNO-A2.0 
catalog\footnote{available at {\tt 
http://archive.eso.org/skycat/servers/usnoa}} as an $R \sim 18.1$, $B \sim 
18.6$ object with identification number U0825\_19648565.
We thus considered it as the optical counterpart of \wga and started
an optical spectrophotometric campaign in order to find out its nature.

\subsection{Spectroscopy}
 
Optical medium-resolution spectra of \wga within the \emph{XMM--Newton}
error circle ($4''$) were acquired on 2003 June 19 with the $4.2$~m
William Herschel Telescope (WHT) located at the Roque de Los Muchachos
Observatory in La Palma, Canary Islands (Spain). The WHT was equipped with
ISIS, which carried on its red arm a 4k$\times$4k MARCONI2 CCD. The spectra
were obtained under a seeing of about $1''$ using Grating R158R, nominally
covering an unvignetted spectral range between 4200 and 8800 \AA. The use
of a $1''$ slit width secured a spectral dispersion of 1.7 \AA/pix. In
total, two $15$--min spectra were acquired between $04:13$ and $04:44$ UT.

Spectra, after correction for flat-field, bias and cosmic-ray rejection,
were background subtracted and optimally extracted (Horne 1986)
using IRAF\footnote{IRAF is the Image
Analysis and Reduction Facility made available to the astronomical
community by the National Optical Astronomy Observatories, which are
operated by AURA, Inc., under contract with the U.S. National Science
Foundation. STSDAS is distributed by the Space Telescope Science
Institute, which is operated by the Association of Universities for
Research in Astronomy (AURA), Inc., under NASA contract NAS 5--26555.}.

Copper--Argon lamps were used for wavelength calibration; the spectra were
then flux-calibrated by using the spectrophotometric standard HD192281
(Massey et al. 1988) and finally averaged together. Correction for slit
losses was also applied to the continuum by checking the flux calibration
against the optical photometry collected in the $UBVRI$ bands (see next
subsection). Wavelength calibration was instead checked by using the
positions of background night sky lines: the error was 0.1 \AA.

Finally, the averaged spectrum was corrected for the foreground Galactic
absorption along the direction of the source assuming a color excess
$E(B-V) = 0.082$~mag, evaluated using the Galactic dust infrared maps by
Schlegel et al. (1998). The reddening value obtained is consistent with the
X-ray absorption column and with the measured interstellar hydrogen column
(Dickey \& Lockman 1990) for an average gas-to-dust ratio (Predehl \& 
Schmitt 1995).

The averaged optical spectrum acquired with WHT (Figure~4) shows a blue
continuum over which a number of emission features are superimposed.
These can readily be identified with typical AGN lines, such as Balmer
lines (H$_\beta$, H$_\gamma$, H$_\delta$), [O~{\sc iii}] $\lambda$5007,
[O~{\sc ii}] $\lambda$3727, and possibly the [Ne~{\sc iii}]
$\lambda$3969/H$_\zeta$ blend, are detected. All these lines point to a
redshift $z = 0.460$ for the source.

Prominent Fe {\sc ii} bumps at around 4500 \AA~and 5200 \AA~(the latter one
falling on the telluric absorption band at 7600 \AA) are also detected at
wavelengths consistent with the above redshift determination.
Table~2 reports full width at half maxima (FWHMs), equivalent
widths (EWs) and fluxes of the detected emission lines.

The FWHM of H$_\beta$ (2000 km s$^{-1}$) and the [O {\sc iii}]/H$\beta$
flux ratio (0.21), measured from the optical spectrum,  are typical of 
NL Type 1 AGN according to the criteria coded by Osterbrock \& Pogge 
(1985).

\begin{figure}[h]
\centering
\label{fig:eww}
\includegraphics[angle=270, scale=0.35]{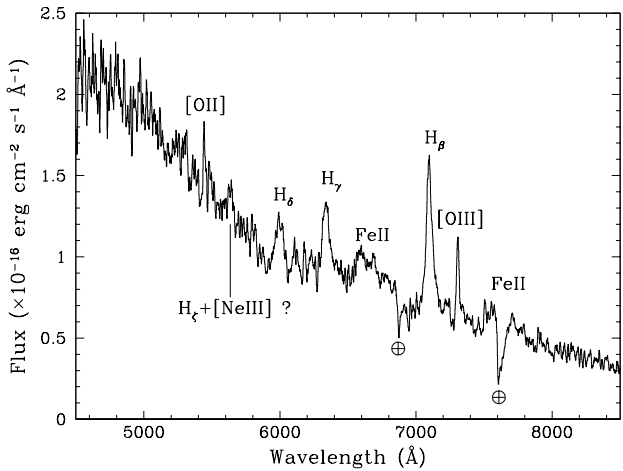}
\caption{Average optical spectrum of \wga obtained with the
WHT at La Palma. The main emission features are labeled. These
allowed us to determine the redshift of the source as $z = 0.460$. The
symbol $\oplus$ indicates atmospheric telluric features.}
\end{figure}

\begin{table}[t!]
\label{optitab}
\caption{FWHMs, EWs and fluxes of the emission lines detected in the
optical spectrum of \wga. Line FWHMs (km~s$^{-1}$) are intrinsic, i.e.
corrected assuming an instrumental broadening of $200$~km~s$^{-1}$; EWs ($\AA$) 
are reported in the source rest frame; flux values are expressed in units of
$10^{-15}$~erg~s$^{-1}$~cm$^{-2}$ and are corrected for the Galactic 
foreground absorption}
\begin{center}
\vspace{-.3cm}
\begin{tabular}{lrrr}
\noalign{\smallskip}
\hline
\noalign{\smallskip}
\multicolumn{1}{c}{Line} & \multicolumn{1}{c}{FWHM} &
\multicolumn{1}{c}{EW} & \multicolumn{1}{c}{Flux} \\
\noalign{\smallskip}
\hline
\noalign{\smallskip}

$[$O {\sc ii}$]$ $\lambda$3727  		& 460$\pm$150 & 4 $\pm$1 & 0.6$\pm$0.2 \\
H$_\delta$                        		& 3300$\pm$500 & 19$\pm$6 & 2.4$\pm$0.7\\
H$_\gamma$                        		& 1700$\pm$400 & 16$\pm$4 & 2.0$\pm$0.5\\
H$_\beta$                        		& 2000$\pm$200 & 46$\pm$8 & 4.7$\pm$0.6\\
$[$O {\sc iii}$]$ $\lambda$5007 		&  700$\pm$100 & 10$\pm$2 & 1.0$\pm$0.2\\

\noalign{\smallskip}
\hline
\noalign{\smallskip}
\end{tabular}
\end{center}
\end{table}

\subsection{Photometry}

Optical photometry was acquired in Loiano (Italy) with the Bologna
Astronomical Observatory $1.52$~meter ``G.D. Cassini'' telescope plus
BFOSC, on 2003 August 1 ($U$ and $R$ bands), 2 ($B$ and $I$ bands) and 3
($V$ band), under an average seeing of 3$''$, 2$''$ and 2$''$,
respectively. The Cassini telescope was equipped with a $1300\times1340$
pixels EEV CCD. This detector, with a scale of 0$\farcs$58/pix, secured a
field of 12$\farcm$6$\times$12$\farcm$6.

Images were corrected for bias and flat-field in the usual fashion and
calibrated using the PG 1633+099 and PG 2213$-$006 fields (Landolt 1992);
the calibration accuracy was better than 3\% in all bands. The source (see
Fig. 1, right panel) was well detected in all optical
filters.  It also showed no significant extension with respect to star-like
objects in the field. Therefore, standard Point Spread Function (PSF)
fitting was chosen as photometry method, and to this aim we used the
{\sl DAOPHOT II} image data analysis package PSF-fitting algorithm (Stetson
1987) running within MIDAS\footnote{MIDAS (Munich Image Data Analysis
System) is developed, distributed and maintained by ESO (European Southern
Observatory) and is available at {\tt
http://www.eso.org/projects/esomidas}}.
\begin{table}[t!]
\label{jou}
\caption{Observed optical--UV--X--ray flux densities of \wga. Optical--UV data
are corrected for Galactic reddening by using $A_{\lambda}/A_V$ given by 
Cardelli et al. (1989) and E(B-V)=0.082. The flux density at 1 keV is 
corrected for the Galactic extinction assuming a column density 
$N_H=5.0\times10^{20}$ cm$^{-2}$} 
\begin{center}
\begin{tabular}{lrr}
\noalign{\smallskip}
\hline
\noalign{\smallskip}
\multicolumn{1}{c}{Instrument} & \multicolumn{1}{c}{$\nu$ (Hz)} & 
\multicolumn{1}{c}{Flux ($\mu$Jy)} \\
\noalign{\smallskip}
\hline
\noalign{\smallskip}

Loiano/UBVRI    & 8.21$\times 10^{14}$& 109$\pm$11 \\
                   & 6.74$\times 10^{14}$& 141$\pm$7  \\
                   & 5.45$\times 10^{14}$& 190$\pm$6  \\
                   & 4.55$\times 10^{14}$& 151$\pm$9  \\
                   & 3.72$\times 10^{14}$& 182$\pm$5  \\ 
\xmm/OM        & 1.33$\times 10^{15}$& 570$\pm$30  \\
                   & 1.48$\times 10^{15}$& 650$\pm$60  \\ 
\xmm/PN        & 2.42$\times 10^{17}$& 0.030$\pm$0.004  \\

\noalign{\smallskip}
\hline

\noalign{\smallskip}
\end{tabular}
\end{center}
\end{table}

Optical fluxes, computed using the tables by Fukugita et al. (1995)
are listed in Table~3, along with those in the UV and X--rays.
Dereddening for optical and UV fluxes was applied assuming the extinction 
law by Cardelli et al. (1989). The X--ray flux density was corrected for the Galactic 
absorption.

\section{Discussion}

Our observations show that the source \wga, located at
about $1.3'$ from the bright radio Galaxy 3C445, is an AGN at redshift
$z=0.46$. It is a quasar as attested by  its optical point-like aspect and  its high luminosity both in the
optical ($M_{\rm R}=-23.2$) and X-ray ($L_{0.2-10~\rm keV}=2.3\times
10^{44}$~erg s$^{-1}$) bands.

More interesting, both the optical and X-ray spectra indicate that 
\wga is a NL QSO. The FWHM of H$_\beta$ (2000 km s$^{-1}$), the [O {\sc iii}]/H$_\beta$
flux ratio (0.21) and the relatively strong FeII emission conform with
the NL Type I classification.
The X-ray spectrum of \wga is very soft as observed in most of the NL Type I AGN
(Boller et al. 1996, Leighly 1999). 
 The steep power law ($\Gamma=2.5$) necessary 
to reproduce the hard X-ray continuum was not sufficient to fit the soft emission 
below 1 keV. PN data need an extra component that, if we choose to model it with a black body 
emission, requires a temperature of  $kT=117$~eV.  Similar black body values characterize the NL QSOs studied by 
 ROSAT and ASCA (Forster \& Halpern 1996, Molthagen et al. 1997, Ulrich et
al. 1999, Komossa et al. 2000, Dewangan et a. 2001).  

Forster \& Halpern (1996) discovered a clear correlation between the soft X-ray
luminosity and the spectral slope in a large sample of NL Type 1 objects
observed with ROSAT.  If we consider a broken power law to model the PN data, the  
fit formally acceptable ($\chi^2=52$ for 49 d.o.f.) yields the following spectral parameters:
$\Gamma_ {\rm soft} = 3.0^{+0.3}_{-0.1}$, $\Gamma_ {\rm hard} =
2.3^{+0.4}_{-0.2}$ with an energy break at $1.1$~keV.
Our source parameters are in perfect agreement with that correlation.
An object with the same intrinsic soft luminosity of \wga $L_{0.1-2~\rm keV} \sim
10^{45}$ erg s$^{-1}$ (assuming $H_0$= 50 and $q_0=0$ to meet the
cosmological assumptions of the authors) is expected to have a soft
spectral slope of about 3, as actually observed.

Finally, the rapid change of the X-ray flux, rather common in ultra-soft AGN (Boller et al.
1996, Leighly 1999), is another hint in favour of its NL classification.

The ultra soft nature of \wga becomes much more evident when the optical UV
and X-ray data are combined to produce a Spectral Energy Distribution
(SED). Unfortunately, neither radio nor infrared counterparts were 
found for this source.

In Fig.~5, the presence of a very prominent UV--soft--X--ray bump is
unequivocal, taking also into account that the X-ray and the UV data are
simultaneous. From Fig. 5, it is also evident that the thermal
model used to parameterize the X-ray excess can not reproduce the entire UV
bump, that is more intense and broad.  On the other hand, it is well known
that a black body emission is a very rough assumption.  Plausible
interpretations assume inverse-Compton scattering of cold disk photons by a
warm/hot atmosphere (Janiuk et al. 2001, Vaughan et al. 2002) or high
accretion slim disk characterized by a multi-colour black body emission
(Mineshinge et al 2000).

\begin{figure}[h]
\label{sed}
\centering
\includegraphics[angle=270, scale=0.35]{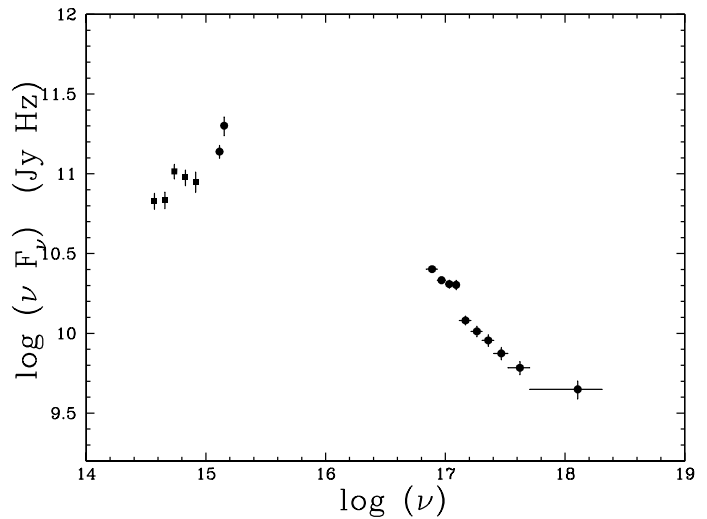}
\caption{Optical-UV-X-ray spectral energy distribution of \wga in the observer frame. A strong UV-soft X-ray bump is clearly visible.}
\end{figure}

A lower limit of the bolometric luminosity $L_{\rm bol} \ge L\sim
3\times10^{45}$ erg s$^{-1}$ can be directly estimated from
Fig.~\ref{sed}, using only the optical--UV--X data, implying a
$0.2-10$~keV X--ray contribute to the total luminosity less than 7$\%$. The
bulk of the emission occurs between $10^{15}$ and $10^{17}$ Hz. By using
simple arguments (uniform accretion at the Eddington limit), we deduce a lower limit to 
the mass of the central supermassive object of 2$\times10^{7}$ M$_{\odot}$.

\begin{acknowledgements}
The authors wish to thank Natalia Auricchio, Ivan Bruni, Elisabetta 
Maiorano and Andrea Simoncelli for the help and assistance with the 
observations made in Loiano. The staff of the Service Observing Programme 
at WHT is gratefully acknowledged. This research has made use of the SIMBAD 
database and of the VizieR catalogue service, both operated at CDS, 
Strasbourg, France. We are also greateful to he ASI for partial support of this research
under grant ASI I/R/069/02.

\end{acknowledgements}

\end{document}